\documentstyle[aps,preprint,tighten,eqsecnum,epsfig,floats,prd]{revtex}

\newcommand{\rf}[4]{{#1} {\bf #2}, #3 (#4)}

\newcommand{\physl}{Phys.\ Lett.\ }

\newcommand{\np}{Nucl.\ Phys.\ }
\newcommand{\beq}{\begin{equation}}
\newcommand{\eeq}{\end{equation}}
\newcommand{\beqa}{\begin{eqnarray}}
\newcommand{\eeqa}{\end{eqnarray}}

\newbox\rotbox

\begin{document}

\preprint{\vbox{
\rightline{ADP-01-45/T477}}}

\draft

\title{Numerical study of lattice index theorem using\\ 
       improved cooling and overlap fermions}

\author{
J.\ B.\ Zhang\footnote{E-mail:~jzhang@physics.adelaide.edu.au},
S.\ O.\ Bilson-Thompson\footnote{
E-mail:~sbilson@physics.adelaide.edu.au},
F.D.R.\ Bonnet\footnote{
E-mail:~fbonnet@physics.adelaide.edu.au},
D.B.\ Leinweber\footnote{E-mail:~dleinweb@physics.adelaide.edu.au 
}, \\
A.G.\ Williams\footnote{E-mail:~awilliam@physics.adelaide.edu.au
} and
J.M.\ Zanotti\footnote{E-mail:~jzanotti@physics.adelaide.edu.au}
}
\address{Special Research Centre for the Subatomic Structure of Matter
and Department of Physics and Mathematical Physics, University of
Adelaide, Adelaide SA 5005, Australia}
\date{\today}

\maketitle

\begin{abstract}
  We investigate topological charge and the index theorem on finite
lattices numerically.  Using mean field improved gauge field
configurations we calculate the topological charge Q using the gluon
field definition with ${\cal O}(a^4)$-improved cooling and an ${\cal
O}(a^4)$-improved field strength tensor $F_{\mu\nu}$.  We also
calculate the index of the massless overlap fermion operator by
directly measuring the differences of the numbers of zero modes with
left- and right--handed chiralities.  For sufficiently 
smooth field configurations we find that the gluon
field definition of the topological charge is integer to better than
1\% and furthermore that this agrees with the index of the overlap
Dirac operator, i.e., the Atiyah-Singer index theorem is satisfied.
This establishes a benchmark for reliability when calculating lattice
quantities which are very sensitive to topology.
\end{abstract}



\section{Introduction}
\label{sec:intro}

  The connection between the topology of a background gauge field and fermion
zero modes is related to the axial anomaly and the large $\eta -\eta^\prime$
mass splitting in QCD.  Lattice gauge theory is the tool best suited to the
study of these nonperturbative issues.  We study this connection in QCD
formulated on a periodic lattice, i.e., on a four-dimensional toroidal
mesh.  For a fine enough lattice and/or a sufficiently smooth gauge
field configuration we should recover the results for continuum QCD on a
4-torus.  In particular, we should recover the Atiyah-Singer index
theorem\cite{as68}.

  In the continuum the Dirac operator, 
$D = \gamma_{\mu}(\partial_{\mu} + igA_{\mu})$,
of massless fermions
in a smooth background gauge field with nontrivial topology
has eigenmodes with zero eigenvalue (i.e., ``zero modes'')
which are also chiral eigenstates of positive or negative
chirality.  The Atiyah-Singer index theorem\cite{as68}
gives the result
\beq
 Q = {\rm index}(D) \, ,
\label{AS_theorem}
\eeq
where 
\beq
Q = \frac{g^2}{32\pi^2}\int d^4x\, \epsilon_{\mu\nu\rho\sigma}
{\rm tr}(F_{\mu\nu}F_{\rho\sigma})
\label{Q_top}
\eeq
is the topological charge of the background gauge field and where 
\beq 
{\rm index}(D) \equiv n_- - n_+
\label{index_D}
\eeq
is the chirality index of the Dirac operator. Here
$ n_+$ and $n_- $ are the number of zero eigenmodes with positive
(right-handed) and negative (left-handed) chiralities respectively, i.e.,
$ D\psi = 0 $ with $+/-$ chiralities such that $ \gamma_5 \psi = \pm \psi$.
These results apply to QCD defined on a continuum 4-torus, where
ultimately we wish to take the size of the 4-torus to infinity, i.e., the
infinite volume limit.

  However, in the lattice formulation one has ambiguities associated
with the discretization and in general one can only expect the index
theorem of Eq.~(\ref{AS_theorem}) to be satisfied on a sufficiently
fine lattice and/or after sufficient smoothing of the gauge fields.
The use of improved operators and actions will lead to the index theorem
being satisfied for less stringent conditions on lattice spacing and/or
smoothness.  There are different ways to calculate $Q$ for a given
gauge field configuration, e.g., using the gauge-field tensor
definition\cite{Pdi81} of Eq.~(\ref{Q_top}) or using the geometrical
definition\cite{lu82}.  In this study we focus on the definition in
Eq.~(\ref{Q_top}).  On a continuum 4-torus the two definitions of topological
charge for the gauge field are necessarily identical.  
However, in the calculation of hadronic observables using typical
lattices, the configurations in the ensembles are too coarse for
lattice definitions of Eq.~(\ref{Q_top}) to lead to integer topological charge and so the
index theorem is not satisfied.

   In an arbitrary gauge field, the Wilson fermion operator does not have
exact zero modes due to the Wilson term which removes fermion doublers
and breaks chiral symmetry.  Attempts to study the index theorem with
such an action requires one to estimate the number of ``zero modes''
by looking at low lying real eigenvalues\cite{SV88}.  Due to the
difficulties in estimating the index of the Wilson fermion operator
in a precise manner, it is difficult to reach a definite conclusion
concerning the validity of the index theorem on a finite lattice with
Wilson fermions.  In the case of Ginsparg-Wilson (GW) Dirac
fermions\cite{GW,has,neub1} there is an exact lattice realization of
chiral symmetry\cite{lus2} and the GW Dirac operator possesses exact zero
modes.  Hence the ambiguity associated with the need to subjectively
estimate the number of ``zero modes'' of the Wilson fermions is absent.  
A good review and introduction to many of these issues can be
found in Ref.~\cite{N98}. 

There are several numerical studies\cite{RP97,CI97,EHN,PH98,TW99,TF00}
of the index theorem on a finite lattice. In Ref.~\cite{RP97} the overlap
fermion formalism is used to estimate the probability distribution of topological
charge $p(Q)$ in pure SU(2) gauge theory by examining the spectral flow of 
$H(\mu)$, where $H(\mu)$ is the hermitian Wilson-Dirac operator. The study in 
Ref.~\cite{CI97} explores the real eigenvalues of the Wilson-Dirac operator, 
which are identified as the lattice counterparts of the continuum zero-modes. They
studied topology by examining the complete spectrum of the fermion matrix in
SU(2) gauge theory on small lattices. Ref.~\cite{EHN} uses the spectral flow
method to perform a comprehensive study of both quenched SU(3) and dynamical 
fermion configurations. There the role of the SW term is also examined.

We focus here on
Neuberger's overlap Dirac operator, which is an explicit solution
of the Ginsparg-Wilson relation, and investigate the Atiyah-Singer
index theorem on a finite lattice by numerical methods. In a recent paper\cite{david},
Adams performed an analytical study which showed that the index theorem was
satisfied by the overlap operator in the continuum limit.
We calculate the index of the overlap fermion operator
directly by measuring the number of left minus right zero modes of
the overlap Dirac operator $D$.

In measuring the gauge-field definition of the topological charge, $Q$,
we use an  ${\cal O}(a^4)$ improved definition of the field strength tensor
leading to an improved topological charge operator.
We also use a mean-field improved Symanzik gluon action in the quenched
approximation to generate our ensemble of configurations.  Where
we have employed cooling to smooth our configurations, we have 
used an  ${\cal O}(a^4)$ improved gluon action in the cooling algorithm.
The resulting $Q$ approaches integer values after a few cooling sweeps
and has been verified to be stable for hundreds to thousands of cooling
sweeps. 

The paper is organized as follows:  In Sec.~\ref{sec:ImpCoolandTopQ}
we introduce the details of our calculation of the improved gauge-field
topological charge $Q$.  In Sec.~\ref{sec:overlap} we review the overlap
quark propagator and describe our calculation of ${\rm index}(D)$.
Our results are described in Sec.~\ref{sec:results} and finally our
summary and discussions are presented in Sec.~\ref{sec:conclusions}.

\section{Improved cooling and Topology}
\label{sec:ImpCoolandTopQ}
To investigate the topology of gauge fields on the lattice we first
construct an ensemble of gauge field configurations.  In lattice QCD,
the gluon fields are represented by $SU(3)$ matrices on each link
connecting adjacent lattice sites.  These links are parallel transport
operators between the lattice sites.  We use a parallel Cabibbo-Marinari\cite{CabMar} 
pseudo-heatbath algorithm with three diagonal $SU(2)$ subgroups looped over twice and
appropriate link partitioning\cite{gfred}.

For typical lattice spacings used in simulations
of QCD the link configurations in the ensemble have fluctuations on many
scales.  In particular, such typical link configurations are not smooth
at the scale of the lattice spacing. However, 
the Boltzmann factor $ e^{-S_{gauge}} $ with $S_{gauge} \propto 1/g^2 $ will ensure that link configurations become
increasingly smooth as the continuum limit is approached ($g \to 0$, $a \to 0$ ). But the cost is that
the volumes that one can afford to simulate at also become correspondingly
smaller.

One common approach to probing the medium to long range topological structure
of typical link configurations is to {\em cool} them sufficiently that they
become approximately smooth at the scale of the lattice spacing\cite{Adel1}.
Cooling involves recursively modifying the link values to locally minimize
the action.  As we sweep repeatedly over the whole lattice we reduce the total 
action, smoothing out fluctuations on successively larger scales\cite{Ring1}. This quickly
eliminates the high-frequency, rough components of the fields, leaving
relatively smooth topological structures.  Once the link configuration has
been cooled sufficiently, we can expect the cooling process to preserve
the global topological charge $Q$.  If we cool indefinitely a link
configuration on a sufficiently fine lattice such that we preserve
the global topological charge $Q$, then we should converge toward
the minimum action solutions with charge $Q$, i.e., we should converge
to self-dual configurations ($F_{\mu\nu}=\tilde F_{\mu\nu}$).
The resulting self-dual configurations on the 4-torus should contain only
$|Q|$ instantons (if $Q>0$) or anti-instantons (if $Q<0$).
The action of an instanton or anti-instanton in infinite space-time is
known analytically and is given by $S_0=8\pi^2/g^2$.  Instantons and
anti-instantons have the property that they are scale invariant, i.e.,
they can have any scale in the infinite 4-volume continuum and their
action $S_0$ is unaffected.  In infinite space-time in the continuum,
a {\em self-dual} gluon configuration with action $S$ will necessarily
have topological charge $|Q|=S/S_0$ and contain only instantons (or
anti-instantons).  To the extent that the finite volume of a
4-torus is sufficiently large compared with the size of the instantons
on it, one can expect the above results to hold .

However, it is well established that cooling with the standard
Wilson action eventually destroys (anti-)instanton configurations, due to
the discretization errors inherent in the Wilson action\cite{Marg93}.
The Wilson gluon (i.e., Yang-Mills) action at each lattice site is
calculated from the plaquette, a closed product of  four link operators
\begin{equation}
S_{\rm Wil} = \beta \sum_{x} \sum_{\mu<\nu}
	\frac{1}{N}Re~{\rm tr}\left(1-U_{\mu \nu}\right)
\label{eq:WilsonAction}
\end{equation}
where the plaquette operator $U_{\mu \nu}$ is 
\begin{equation}
U_{\mu \nu} = U_{\mu}(x)U_{\nu}(x+\mu)U^{\dag}_{\mu}(x+\nu)
  U^{\dag}_{\nu}(x) \, .
\label{eq:plaquette}
\end{equation}
This Wilson plaquette action contains deviations from the continuum Yang-Mills
action of ${\cal O}(a^2)$ where $a$ is the lattice spacing. This problem
may be remedied by improving the  action.  Tree-level improvement of the
classical lattice action combined with mean-field (i.e., tadpole)
nonperturbative improvement \cite{LePage} provide a simple and very effective 
means of eliminating lowest-order discretization errors.  The simplest
improvement of the classical gauge action on the lattice is achieved
by taking a linear combination of the Wilson plaquette and an 
$a \times 2a$ rectangle, i.e., Symanzik improvement.  Since the plaquette
and rectangle have different ${\cal O}(a^2)$ errors they can be added in 
a linear combination in such a way that these ${\cal O}(a^2)$ errors
cancel leaving only errors of ${\cal O}(a^4)$ in the classical
(i.e., tree-level) gauge action.  One can preserve this improvement
at the nonperturbative level by combining this with mean-field improvement
of the links\cite{Lepa93}.  More recently DeForcrand 
{\em et al.} \cite{DeForc} have used tree-level improvement to construct a
lattice action which eliminates ${\cal O}(a^4)$ errors and leaves only
${\cal O}(a^6)$ errors, by using combinations of up to five different
closed-loop products of link operators (Wilson loops).  From these five
loops,  particular linear combinations were studied in detail and were
referred to as 3-loop, 4-loop, and 5-loop improved actions.  The
difference is the number of non-zero contributions in the linear
combination of the five planar loops.  They represent the
improved action as
\beq
S_{\rm Imp} = \sum_{i=1}^{5} c_i S_i~,
\label{eq:DeForcAction}
\eeq
where the $S_i$ are the actions calculated as per equation
(\ref{eq:WilsonAction}), but using five different Wilson loops, and
the $c_i$ are the weighting constants tuned to eliminate the
${\cal O}(a^2)$ and ${\cal O}(a^4)$ errors. Additional details may be found in Ref.\cite{DeForc}. 

In our work we construct improved actions utilizing the results of
DeForcrand {\em et al.} However we have chosen to improve the topological charge  
via an improved field strength tensor.
In particular, we employ an ${\cal O}(a^4)$ improved definition of $F_{\mu\nu}$ in
which the standard clover-sum of four $1 \times 1$ Wilson loops lying in the $\mu ,\nu$
plane is combined with $2 \times 2$ and $3 \times 3$ Wilson loop clovers.
Bilson-Thompson {\it et al.} \cite{sbilson} find

\beq
gF_{\mu\nu} = {-i\over{8}} \left[\left( {3\over{2}}W^{1 \times 1}-{3\over{20u_0^4}}W^{2 \times 2}
+{1\over{90u_0^8}}W^{3 \times 3}\right) - {\rm h.c.}\right]_{\rm Traceless}~,
\eeq
where $W^{n \times n}$ is the clover-sum of four $n \times n$ Wilson loops and where
$F_{\mu\nu}$ is made traceless by subtracting $1/3$ of the trace from each diagonal
element of the $3 \times 3$ color matrix. This definition reproduces the continuum limit
with ${\cal O}(a^6)$ errors. We employ the plaquette measure of the mean link
\beq
 u_0 = \langle \frac{1}{3} Re~{\rm tr} U_{\mu\nu}\rangle^{1/4}_{x,\mu\ne \nu}~,
\eeq
which is updated after each sweep through the lattice. The mean link $u_0$ rapidly tends to one under 
cooling, thus reproducing the classical limit. However, early in the cooling procedure,
the fields are not classical and $u_0$ serves to tadpole improve the definition of  $F_{\mu\nu}$.

This improved field-strength tensor can be used directly in
Eq.~(\ref{Q_top}) resulting in a topological charge which
is automatically free of discretization errors to the same order as
the field-strength tensor. On self-dual configurations, this operator produces integer
topological charge to better than 4 parts in $10^4$.
 Furthermore, since the gluon (i.e., Yang-Mills)
action is also based upon the field-strength tensor, it is possible to
create what we refer to as a {\em reconstructed action} based upon the
improved field-strength tensor.  The value of the action calculated with
the reconstructed action operator can be compared with the value calculated
with the standard improved action operator Eq.~(\ref{eq:DeForcAction}) 
at each cooling sweep as a double-checking mechanism to ensure that the
two different approaches to tree-level improvement removing 
${\cal O}(a^4)$ errors yield consistent results.  

\begin{figure}[t]
\begin{center}
\epsfig{figure=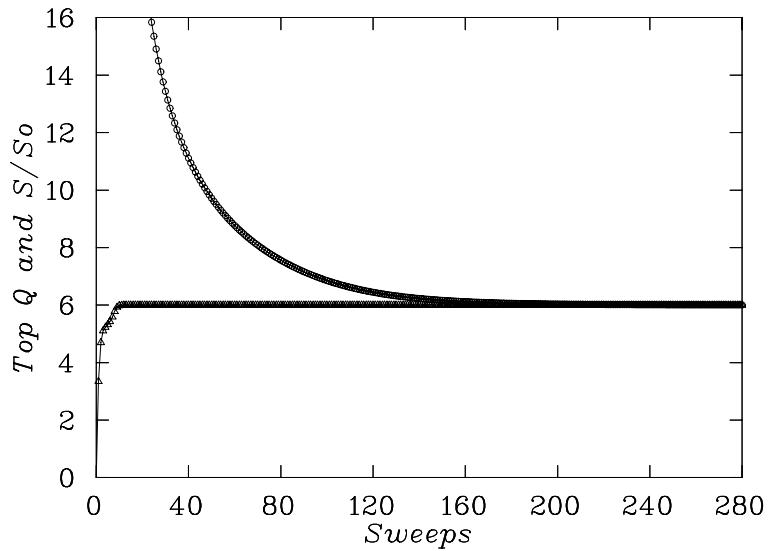,angle=90,width=10cm}
\end{center}
\caption{An example plot of how improved cooling stabilizes the action (circles) and 
topological charge (triangles) at consistent values. The action has been rescaled by dividing it
by $S_0$, the action of a single instanton.}
\label{fig:cooling}
\end{figure}

The principal criteria by which we may judge the value of an
improvement scheme are how quickly and how closely the results for the
topological charge $Q$ approach integer values, and the stability with 
which they remain at that integer. On a lattice, the
discretization errors prevent us from obtaining exactly integer results.
However the more successful the improvement program, then the more
rapidly we can expect cooling to lead us to a stable $Q$ and the closer
this will be to an integer.  Figure \ref{fig:cooling} shows how 
both the action and topological charge approach the same integer value
as a function of the number of cooling sweeps.  This is exactly
as we would expect for $Q$. As we approach
self-duality on the 4-torus we appear to recover $S/S_0=|Q|$ just as
in the continuum infinite-volume limit. 
Recall that the positivity of $ (F_{\mu \nu}^a \pm \tilde{F}_{\mu \nu}^a)^2 $ ensures that 
$ 0 \leq \int d^4 x(F_{\mu \nu}^a \pm \tilde{F}_{\mu \nu}^a)^2 = 2 \int d^4 x [
(F_{\mu \nu}^a)^2 \pm F_{\mu \nu}^a \tilde{F}_{\mu \nu}^a] $ and hence 
\beq
S = \frac{1}{4} \int d^4 x (F_{\mu \nu}^a)^2 \geq \frac{1}{4 } \mid \int d^4 x 
F_{\mu \nu}^a \tilde{F}_{\mu \nu}^a \mid = \frac{8 \pi^2}{g^2}\mid Q \mid = S_0 \mid Q \mid ,
\eeq
where we have defined $ S_0 \equiv {8 \pi^2}/{g^2}$. Self-dual configurations 
$ F_{\mu \nu}^a = \pm \tilde{F}_{\mu \nu}^a $ saturate this identity, i.e. , $\mid Q \mid$ = $S/S_0$. 
This result applies independent of the shape of the space-time manifold and so applies on the 
continuum 4-torus as well as for infinite space-time.  However, note that Nahm's theorem\cite{nahm}
implies that there is no self-dual $\mid Q \mid = 1 $ configuration possible on the 4-torus. 
In infinite space-time $S_0$ is the single instanton or anti-instanton action. 

While an in-depth comparison of the various cooling schemes is beyond
the scope of this report, we summarize by noting that it has been
determined\cite{sbilson} that 3-loop improved cooling with a 3-loop
topological charge operator gives excellent results in terms of how
close the calculated values of $S/S_0$ and $Q$ come to integer values,
the stability with which they remain at that integer, and in terms of
the speed with which cooling can be performed.

\section{Overlap fermions and zero modes}
\label{sec:overlap}

 The overlap fermion formalism\cite{neub2} provides a way of realizing 
exact chiral symmetry on the lattice.  The massless overlap-Dirac operator
can be written as
\beq
   D_{\rm o}(0) = ( 1 + \gamma_5 \epsilon(H_{\rm w})) \, ,
\eeq
where $\epsilon(H_{\rm w})$ is the matrix sign function
\beq
   \epsilon(H_{\rm w}) = \frac{H_{\rm w}}{|H_{\rm w}|} \, ,
~~~~~H_{\rm w} = \gamma_5 D_{\rm w},
\eeq
and $ D_{\rm w} $ is the usual Wilson-Dirac Operator
\beqa
D_{\rm w}  &= &M_{a\alpha m, b\beta n} \nonumber   \\
&=&\delta_{a,b} \delta_{\alpha,\beta} \delta_{m,n}
-\kappa \sum_{\mu=1}^4 [({\bf 1}-\gamma_\mu)_{\alpha\beta} U_\mu(m)_{ab}
\delta_{m,n-\hat{\mu}}
+({\bf 1}+\gamma_\mu)_{\alpha\beta} {U_\mu^\dagger(m-\hat{\mu})_{ab}}
\delta_{m,n+\hat{\mu}}]\\
\eeqa
and where $ \kappa $ is the hopping parameter,
\beq
\kappa =\frac{1}{-2m+8} \, .
\eeq
The bare mass parameter for the Wilson kernel in the overlap formalism
$m$ has to be in the range (0,2) for $D_{\rm o}(0)$ to describe
a single massless Dirac fermion.   Note that the $m$ used in this context
is the negative of the bare mass used in simulations with the Wilson quark
action itself.  In principle, any value of $m$ in the above range 
should give the same continuum theory.  But on a finite lattice, where
volume and lattice spacing are finite, the results for the overlap action
can depend on the the value chosen for $m$.   For the 
purposes of using overlap fermions to be most sensitive to the topology
of the background gauge fields\cite{EHN}, 
$m$ has to be chosen $ m > m_1(g^2)$ for some $m_1(g^2)$ going to zero
in the continuum limit. For typical gauge configurations $ m_1(g^2)$
is slightly less than $ m_c$, where $m_c$ is the critical value of $m$
at which the pion mass extrapolates to zero in a simulation with
ordinary Wilson fermions. In the hopping
parameter formalism, the $ \kappa $ should to be in the range
($\kappa_c$, 0.25) as tree level.  The massless overlap operator $D_{\rm o}(0)$
has been shown\cite{neub1} to satisfy the Ginsparg-Wilson
relation
\beq
\{\gamma_5, D_{\rm o}(0)\} = D_{\rm o}(0)\gamma_5 D_{\rm o}(0).
\eeq
Its spectral properties are
\begin{itemize}
	\item The modulus of the eigenvalue of $D_{\rm o}(0)$ lies in the range [0, 2].
	It has exact zero eigenvalues which are associated with topology.
	The corresponding eigenvectors are also eigenvectors of $\gamma_5$.
        They need not occur in pairs.
	There can be $n_+$ zero-modes with eigenvalues of $\gamma_5$ equal
        to 1
	and $n_-$ zeromodes with eigenvalues of $\gamma_5$ equal to -1.
        It also has eigenvalues equal to
	$\pm 2$ which also have definite chirality. Their difference
	$(n^{\prime}_+ - n^{\prime}_-)$ is equal to $(n_- - n_+)$.

	\item Non-zero eigenvalues of  $D_{\rm o}(0)$ are complex, their
        moduli are less
	than 2, they come as pairs and are conjugate to each other. The
        associated
	eigenvectors are not eigenvectors of $\gamma_5$.
\end{itemize}

In practice, we calculate a small number of low-lying eigenvalues and
eigenvectors
of $ D_{\rm o}^\dagger(0)D_{\rm o}(0) $. Note that $ D_{\rm o}^\dagger(0)
D_{\rm o}(0) $ commutes with
$D_{\rm o}(0)$ and it is hermitian and positive definite. It also commutes 
with
$\gamma_5$ and can be simultaneously diagonalized. Hence, $ D_{\rm o}^\dagger(0)
D_{\rm o}(0) $
has zero eigenmodes of definite chirality which are also the
zero eigenmodes of $D_{\rm o}(0)$.

 We compute low-lying eigenmodes of $ D_{\rm o}^\dagger(0)D_{\rm o}(0) $
using the Ritz functional algorithm \cite{TH96}. In the computation of
the overlap operator, the time consuming part is the calculation of the matrix
sign function $\epsilon (H_{\rm w})$. There are several numerical
approaches by which  to approximate the sign function
$\epsilon (z)$~\cite{neu98c,ehn99a,hjl99}.
We adopt the optimal rational approximation~\cite{ehn99a} with a ratio of
polynomials of degree 12 in the Remez algorithm. We find the error
to the approximation of $\epsilon (z)$
to be within $10^{-6}$ in the range [0.04, 1.5] of the argument $z$.
To improve the accuracy as well as efficiency of computation of the matrix
sign function $\epsilon (H_{\rm w})$, a number of low-lying eigenvalues of
$H_{\rm w}$
whose absolute value are less than 0.04 are projected out. Since
\beq
D_{\rm o}^\dagger(0)D_{\rm o}(0) = D_{\rm o}^\dagger(0) + D_{\rm o}(0),
~~~~D_{\rm o}^\dagger(0)=\gamma_5 D_{\rm o}(0)\gamma_5,
\eeq
we use chiral states in the Ritz functional algorithm, and hence can save
one matrix multiplication~\cite{ehn99b} per iteration.

\section{Results and comparison}
\label{sec:results}

Configurations have been generated on an $8^3\times{16}$ lattice at both
$\beta=4.80$ and $\beta=4.38$ as well as on a $12^3\times{24}$ lattice at
$\beta=4.60$.
Configurations are selected after $N_{\rm therm}=5000$ thermalization sweeps
from a cold start and every $N_{\rm sep}=500$ sweeps thereafter with the
average link $u_0$ fixed at the time of the first sample configuration
being taken.  Lattice parameters are summarized in
Table~\ref{simultab}.

\begin{table}[b]
\caption{Parameters of generated lattices.}
\begin{tabular}{cccccccc}
Action &Volume &$N_{\rm{therm}}$ & $N_{\rm{sep}}$ &$\beta$ &$a$ (fm) & $u_{0}$ & Physical Volume (fm)\\
\hline
Improved       & $8^3\times{16}$ & 5000 & 500 & 4.80 & 0.093  & 0.89625 & $0.75^3\times{1.50}$ \\
Improved       & $8^3\times{16}$ & 5000 & 500 & 4.38 & 0.165  & 0.87821 & $1.32^3\times{2.64}$ \\
Improved       & $12^3\times{24}$ & 5000 & 500 & 4.60 & 0.125  & 0.88888 & $1.50^3\times{3.0}$ \\
\hline
\end{tabular}
\label{simultab}
\end{table}

We first use the gluon field definition to calculate the topological charge
$Q$ by using the three-loop improved field strength $F_{\mu\nu}$ as described
in Sec.~\ref{sec:ImpCoolandTopQ}.  For each configuration we measure
the topological charge by cooling with the 3-loop improved action for
just enough cooling sweeps that $Q$ is integer to within 1\%.
This typically requires from 1-30 cooling sweeps depending on the
lattice spacing and the particular configuration.  We retain both the
original uncooled configurations as well as these ``just-cooled''
configurations. We denote the topological
charge obtained from the ``just-cooled'' gluon field configurations
as $Q_{\rm g}$. 

Secondly, we use the overlap formalism to
calculate the index of the gauge configurations at
two $\kappa$ values, i.e.,  $\kappa = 0.19 $ corresponding 
to $ m = 1.36$  and $\kappa = 0.2499 $ corresponding to 
$m = 1.999$.  The second value approaches the largest allowed value of $m$
to describe a single massless Dirac fermion. 
We extract the overlap index for both the original uncooled and the
``just-cooled'' cooled configurations and denote the results by
$Q_{\rm d}$ and $Q_{\rm d-cooled}$ respectively.

\begin{table}[tp]
\caption{Results for $\beta=4.80$ improved gauge configurations on an 
$8^3 \times 16$ lattice with spacing $a = 0.093$ fm:
$Q_{\rm d}$ is calculated from the zero-modes of the overlap operator on
the original uncooled configurations; $Q_{\rm g}$ is obtained
using the improved topological charge operator for the ``just-cooled''
configurations, (i.e., configurations after just-enough improved
cooling sweeps to bring $Q$ within 1\% of integer);
$Q_{\rm d-cool}$ is also calculated from the zero-modes
of the overlap operator, but on the ``just-cooled'' configurations.}
\begin{tabular}{ccccc}
Configuration \# & $Q_{\rm d}(\kappa = 0.19)$&$Q_{\rm d}(\kappa = 0.2499)$ 
     & $Q_{\rm g}$ &$Q_{\rm d-cool}$ \\
\hline
1   &   0&   0   &   0   &   0    \\
2   &   0&   0   &   0   &   0    \\
3   &   0&   0   &   0   &   0    \\
4   &  +1&  +1   &  +1   &  +1    \\
5   &   0&   0   &   0   &   0    \\
6   &  $-1$&  $-1$   &  $-1$   & $-1$    \\
7   &   0&   0   &   0   &   0    \\
8   &   0&   0   &   0   &   0    \\
9   &   0&   0   &   0   &   0    \\
10  &   0&   0   &   0   &   0    \\
11  &   0&   0   &   0   &   0    \\
12  &   0&   0   &   0   &   0    \\
13  &   0&   0   &   0   &   0    \\
14  &   0&   0   &   0   &   0    \\
15  &   0&   0   &   0   &   0    \\
16  &   0&   0   &   0   &   0    \\
17  &   0&   0   &   0   &   0    \\
18  &   0&   0   &   0   &   0    \\
19  &   0&   0   &   0   &   0    \\
20  &   0&   0   &   0   &   0    \\
21  &   0&   0   &   0   &   0    \\
22  &   0&   0   &   0   &   0    \\
23  &   0&   0   &   0   &   0    \\
24  &  $-1$&  $-1$   &  $-1$   &  $-1$    \\
25  &  $-2$&  $-2$   &  $-2$   &  $-2$    \\
26  &  +1&  +1   &   0   &   0    \\
27  &   0&   0   &   0   &   0    \\
28  &   0&   0   &   0   &   0    \\
29  &   0&   0   &   0   &   0    \\
30  &   0&   0   &   0   &   0    \\
\hline
\end{tabular}
\label{resb480}
\end{table}

The results for the small-volume $8^3\times 16$, $\beta=4.80$ lattice
are collected in Table~\ref{resb480}.  We see that since
the physical volume of our lattice is rather small, only a few
configurations have non-trivial topology. The index $Q_{\rm d}$
of the overlap operator calculated at the two different $\kappa $ values
are the same for all 30 configurations.  The range of cooling sweeps
needed to get $Q$ to within 1\% of integer was between 1 and 7 with
an average of approximately 3.  The index for these ``just-cooled''
configurations differs from these for only one of the thirty
configurations, i.e., configuration 26.  For all of the configurations
the gluon topological charge $Q_{\rm g}$ is identical to the  
the index of the overlap operator extracted from ``just-cooled''
configurations $Q_{\rm d-cooled}$. 
These results indicate that on such a fine lattice ($a=0.093$~fm)
topology is relatively well represented and the index theorem is
``almost'' valid for uncooled configurations.  We see also that for
``just-cooled'' configurations the index theorem appears to be
perfectly satisfied. Future investigations should explore the possible 
volume dependence of these results. 

\begin{table}[t]
\caption{Results for $\beta=4.38$ improved gauge configurations on an $8^3 \times 16$ lattice with 
spacing $a = 0.165$~fm:
$Q_{\rm d}$ is calculated from the zero-modes of the overlap operator on
the original uncooled configurations; $Q_{\rm g}$ is obtained
using the improved topological charge operator for the ``just-cooled''
configurations;
$Q_{\rm g3}$ is obtained using the improved topological charge operator
for the configurations obtained after just 3 improved cooling sweeps;
$Q_{\rm d-cool}$ is also calculated from the zero-modes
of the overlap operator, but on the ``just-cooled'' configurations.
}
\begin{tabular}{cccccc}
Configuration \# & $Q_{\rm d}(\kappa = 0.19)$&$Q_{\rm d}(\kappa = 0.2499)$ 
& $Q_{\rm g3}$&$Q_{\rm g}$ &$Q_{\rm d-cool}$ \\
\hline
1   &   0   &   0  &  +1 &   0   &   0    \\
2   &   +3  &  +3  &  +4 &  +3   &  +3    \\
3   &   +1  &   0  &   0 &   0   &   0    \\
4   &    0  &   0  &   0 &   0   &   0    \\
5   &    0  &   0  &  $-1$ &  $-1$   &  $-1$    \\
6   &   $-1$  &  $-1$  &  $-1$ &  $-1$   &  $-1$    \\
7   &   +5  &  +4  &  +4 &  +6   &  +6    \\
8   &   $-1$  &  $-4$  &  $-2$ &  $-2$   &  $-2$    \\
9   &   +1  &  +1  &  +1 &  +1   &  +1    \\
10  &    0  &   0  &  $-1$ &   0   &   0    \\
11  &   $-2$  &  $-2$  &  $-3$ &  $-3$   &  $-3$    \\
12  &   $-1$  &  $-1$  &  $-1$ &  $-1$   &  $-1$    \\
13  &   +2  &  +2  &  +3 &  +2   &  +2    \\
14  &   +3  &  $-5$  &  $-4$ &  $-4$   &  $-4$    \\
15  &   $-3$  &  $-3$  &  $-3$ &  $-3$   &  $-3$    \\
16  &    0  &   0  &  $-1$ &   0   &   0    \\
17  &    0  &  $-1$  &  $-1$ &   0   &   0    \\
18  &   $-1$  &  $-1$  &  $-1$ &  $-1$   &  $-1$    \\
19  &   +1  &  +1  &  +1 &  +1   &  +1    \\
20  &   +1  &   0  &   0 &  $-2$   &  $-2$    \\
\hline
\end{tabular}
\label{resb438}
\end{table}

We next turn to the coarser ($a=0.165$~fm), larger-volume $8^3\times 16$,
$\beta=4.38$ lattice.  The results for this lattice are shown in
Table~\ref{resb438}.  Since the physical volume of the lattice is larger
there is more nontrivial topology than before, which is reflected in the large
$Q$ values.  We find that $Q_{\rm d}$ now differs in 30\% of cases depending
on which value of $\kappa$ is used in the overlap kernel.  In some cases,
e.g., for configuration 14, the disagreement is very significant indeed.
The range of cooling sweeps needed for this coarse lattice
was between 4 and 28 with an average of approximately 12.5.
In addition to measuring the gluon topological charge on these ``just-cooled''
configurations $Q_{\rm g}$ we have on this lattice also attempted to
identify the gluon topological charge after just three cooling sweeps, i.e.,
$Q_{\rm g3}$.  This was done by identifying the nearest integer to
the gluon topological charge after just three cooling sweeps and 
was motivated by the observation that the action density
and the reconstructed action density matched within 10\% after three sweeps.
We see that $Q_{\rm g3}$ is reasonably consistent with the robust
$Q_{\rm g}$, but there are significant differences.  $Q_{\rm g3}$ also
has significant differences from the $Q_d$ values,
The disagreement between $Q_{\rm d}$ and $Q_{\rm g}$ are more significant
on this coarse lattice.

 These differences may be interpreted by considering the size of the topological
objects giving rise to exact zero modes\cite{Edku}. At $\kappa = 0.19 $, the overlap
operator will typically to miss zero modes associated with small topological objects. Indeed,
for the six configurations where $Q_{\rm d}(\kappa = 0.19 )$ $\ne$ $Q_{\rm d}(\kappa = 0.2499 )$,
$Q_{\rm d}(\kappa = 0.2499 )$ agrees better with $Q_{\rm g3}$ than $Q_{\rm g}$. This
is as one might expect, as further cooling will remove topological objects smaller than the 
dislocation threshold of the improved cooling algorithm, which is typically two lattice spacings. 

These results all suggest that for uncooled configurations on this
coarser lattice topology is not well represented and the index theorem
is badly violated.  However, from the perfect agreement between
$Q_{\rm g}$ and $Q_{\rm d-cool}$, we see that for ``just-cooled''
configurations the index theorem is again perfectly satisfied.  In
that sense, we conclude that our ``just-cooled'' configurations are
indeed smooth enough.

\begin{table}[tp]
\caption{Results for $\beta=4.60$ improved gauge configurations on an $12^3 \times 24$ lattice with 
spacing $a = 0.125$~fm:
$Q_{\rm d}$ is calculated from the zero-modes of the overlap operator on
the original uncooled configurations; $Q_{\rm g}$ is obtained
using the improved topological charge operator for the ``just-cooled''
configurations;
$Q_{\rm d-cool}$ is also calculated from the zero-modes
of the overlap operator, but on the ``just-cooled'' configurations.
}
\begin{tabular}{ccccc}
Configuration \# & $Q_{\rm d}(\kappa = 0.19)$ & $Q_{\rm g}$ 
       &$Q_{\rm d-cool}$ \\
\hline
1   &   $-1$   &  $-1$   &  $-1$    \\
2   &   +1   &  +1   &  +1    \\
3   &   $-4$   &  $-4$   &  $-4$    \\
4   &    0   &   0   &   0    \\
5   &   +1   &  +1   &  +1    \\
6   &   $-2$   &  $-2$   &  $-2$    \\
7   &   $-2$   &  $-2$   &  $-2$    \\
8   &   $-1$   &  $-1$   &  $-1$    \\
9   &    0   &   0   &   0    \\
10  &   $-2$   &  $-2$   &  $-2$    \\
11  &   +1   &  +1   &  +1    \\
12  &    0   &   0   &   0    \\
13  &   $-3$   &  $-3$   &  $-3$    \\
14  &   $-4$   &  $-3$   &  $-3$    \\
15  &   $-2$   &  $-4$   &  $-4$    \\
16  &   $-1$   &  $-2$   &  $-2$    \\
17  &   $-5$   &  $-5$   &  $-5$    \\
18  &   $-2$   &  $-3$   &  $-3$    \\
19  &   $-5$   &  $-5$   &  $-5$    \\
20  &   $-4$   &  $-4$   &  $-4$    \\
\hline
\end{tabular}
\label{resb460}
\end{table}

   We now present in Table~\ref{resb460} the results for the third lattice
with an intermediate lattice spacing $a=0.125$~fm
corresponding to $\beta = 4.60$ and
with lattice size $12^3\times 24$.  This lattice has the largest physical
volume.  The range of cooling sweeps needed for this medium-spaced lattice
was between 2 and 14 with an average of approximately 8.
As we might anticipate, the agreement of $Q_{\rm g}$ with $Q_{\rm d}$
in this case is worse than for the fine lattice but better than for the
coarse lattice.  We do not have results for the large
$\kappa=0.2499$ case here, since on this larger lattice this marginal
choice of $\kappa$ proved numerically difficult.
Calculating $Q_{\rm d-cool}$ with its ``just-cooled'' configurations
gives perfect agreement with $Q_{\rm g}$ for all configurations.

It is expected that $\langle Q^2\rangle$ should scale approximately
as the volume $V$ for large volumes, since the topological
susceptibility is given by $\langle Q^2\rangle/V$ and should be volume
independent.  The ratio of this 4-volume to that for the
coarse lattice is 1.66.  The mean $Q^2$ per configuration for this
lattice is easily seen from Table~\ref{resb460} to be
$\langle Q^2\rangle=7.3$, whereas for the coarser, smaller-volume lattice in
Table~\ref{resb438} we find $\langle Q^2\rangle=4.85$.
We see that
$\langle Q^2\rangle_{\rm big}/\langle Q^2\rangle_{\rm small}=1.51$
which is approximately equal to $V_{\rm big}/V_{\rm small}=1.66$.
This level of agreement seems reasonable for such modest volume
lattices and small numbers of
configurations.

We also observe that for this largest lattice there
appears to be a significant imbalance in the sign of $Q$ in the ensemble
and in particular the sign is consistently negative for the last 8
configurations in the ensemble.  If the configurations
were uncorrelated, this would be very unlikely to occur.
This suggests that for larger lattices,
when measuring quantities that are very sensitive to topology, one should
use an increased number of thermalization and separation sweeps.  To
balance the topological charge in the ensemble one should perhaps
consider doubling the ensemble size by adding parity-transformed
link configurations\cite{derek} which leave the action invariant but reverse
the sign of $Q$.

\section{Conclusions and Discussions}
\label{sec:conclusions}

   We have shown that overlap fermions are suitable for use in the 
study of topology and the Atiyah-Singer index theorem in lattice
simulations.  We have shown that with an improved gluon action and
an improved definition of the gluon topological charge operator we can
obtain near integer topological charge (within 1\%) within a relatively
small number of cooling sweeps.  The finer the lattice the fewer cooling
sweeps are needed to obtain these ``just-cooled'' configurations.
We found that for lattice spacings of $a=0.093$, 0.125 and 0.165~fm
we needed on average 3, 8, and 12.5 improved cooling sweeps respectively.
For all configurations on all lattices, we found the index theorem satisfied
for these ``just-cooled'' configurations.  Even for the finest lattice
and with an improved topological charge operator some small number of
cooling sweeps was needed to obtain an integer gluon topological charge.
On the finest lattice the index of the overlap operator appeared
independent of $\kappa$ and agreed with the ``just-cooled'' overlap
index 29 out of 30 times.  This lattice appears then to be almost fine
enough that the index theorem has meaning without cooling.  For
all lattices the index theorem appeared to be fully satisfied for
the just-cooled configurations.  This provides us with a clear
benchmark for smoothness and lattice spacing when calculating
lattice quantities which are very sensitive to topology.

\section*{Acknowledgments}

We thank Paul Coddington of the Distributed and High-Performance Computing
Group and Francis Vaughan of the South Australian Centre for
Parallel Computing for support in the development of parallel algorithms in
High Performance Fortran (HPF).  This work was carried out using the
Orion Supercomputer.  The support of the Australian Research Council for
this research is gratefully acknowledged.


\end{document}